\documentclass[twocolumn,showpacs,preprintnumbers]{revtex4}

\usepackage[dvips]{graphicx}
\usepackage{dcolumn}
\usepackage{bm}
\begin{document}
\title{Tracer density discontinuities in turbulent flows: simple model and scaling laws}
\author{Jaan Kalda and Aleksandr Morozenko}
\affiliation{
Institute of Cybernetics,
Tallinn Technical University,
Akadeemia tee 21,
12618 Tallinn,
Estonia
}
\begin {abstract} 
Mixing in fully developed incompressible turbulent flows is known to lead to a cascade of discontinuity fronts of passive scalar fields.
A one-dimensional (1D) variant of Baker's map is developed, capturing the main mechanism responsible for  the 
emergence of these discontinuities. For this 1D model, expressions for the height-distribution function of the 
discontinuity fronts and structure function scaling exponents $\zeta_p$ are derived [for Kolmogorov turbulence, $\zeta_p=\frac 23\log_3(p+1)$]. These analytic findings are 
in a good agreement with both our 1D simulations, and the results of earlier numerical and experimental studies. 
\end {abstract} 
\maketitle

Explaining the origin of intermittency 
in turbulent media is a long-standing challenge for statistical physicists; a particular attention has been paid to the anomalous behavior of the structure functions scaling exponents $\zeta_p$.
While the first studies in this field date back to almost five decades \cite{K62}, 
the theory is  still far from being complete. 
Even in the simplest case of passive scalar turbulence, despite of extensive studies
(c.f.\ reviews \cite{Falk,reviews}), 
the theoretical understanding of the anomalous scaling  is rather sketchy: the exponents are either obtained experimentally or numerically (e.g.\ \cite{Antonia,Meneveau,Ruiz,Mydlarski,Chen,Celani,Moisy}); 
the analytic results are limited to specific velocity spectra (not applicable to the Kolmogorov case) \cite{Falk,Frisch}. There are also 
hierarchical models \cite{phenomenol} (derived from the velocity field intermittency model \cite{She}), 
which can fit the experimental data relatively well (although being inconsistent with the Obukhov-Corrsin result for $\zeta_2$ \cite{Obukhov}), but rely on a couple of phenomenological hypothesis;
therefore, their usefulness in understanding the origins of intermittency is limited.

It is known that if a passive scalar evolves in fully developed turbulent flows,
the scalar field becomes everywhere discontinuous: the discontinuity fronts  of fractal structure will emerge \cite{Celani}.
These fronts are the very reason for the anomalous scaling of structure functions. 
However, little is known about the formation and  statistics  of them.

In the first part of the Letter, we outline qualitatively the mechanism of the formation of passive scalar discontinuity fronts, 
and construct a 1D model incorporating all the essential features of that mechanism. 
In the second part, we analyze the properties of our  model theoretically, applying a non-rigorous scaling analysis, present the simulation results, and 
compare them (as well as some earlier experimental and numerical results) with the  theoretical scaling laws.

{\em I.\ Formation of discontinuities.} 
In what follows, we assume that  the tracer density $\theta(r,t)$ evolution is described by a simple diffusion equation,
\begin{equation} \label{ode}
\partial_t \theta + \bm{v}\bm{\nabla}\theta=\kappa \bm \nabla^2 \theta +f(\bm{r}, t),
\end{equation}
where $\bm{v}(\bm{r},t)$
is a turbulent velocity field, $f(\bm{r}, t)$ is a forcing, and the seed diffusivity $\kappa$ is assumed to be very small (but not zero).
Finally, it is assumed that the velocity 
field obeys a power-law Kraichnan statistics,
\begin{equation}\label{cf2}
\begin{array}{l}
\left<v_i(\bm{r},t)v_j(\bm{r}^\prime,t^\prime)\right> = 2\delta(t-t^\prime)D_{ij}(\bm r - \bm r^\prime),\\
d_{ij}(\bm r)=D_1[(d-1+\xi)\delta_{ij}r^\xi-\xi r_ir_jr^{\xi-2}],
\end{array}
\end{equation}
where $d_{ij}(\bm r)$ is the non-constant part of $D_{ij}(\bm r)$,
$D_1$ --- a constant, and $\xi$ --- the smoothness exponent, c.f.\ \cite{Falk} and references therein.
So, we neglect the intermittency of the underlying velocity field. It should be noted that
according to the experimental and numerical evidence, the passive scalar intermittency is stronger than
the velocity field intermittency (e.g.\ characterized by greater anomality $2\zeta_2-\zeta_4$), c.f.\ \cite{Xu}. Thus, one can expect that 
the former dominates over the latter, and the behavior of tracers in real (intermittent) velocity fields is very similar to that of 
in idealized Gaussian fields (for a numerical evidence, c.f.\ \cite{Zhao}). Finally, note that owing to the robustness of our model, the assumption of delta-correlation in time will not be actually used;
Eq.~(\ref{cf2}) is adopted for a starting point merely to simplify comparisons with other studies.

Mixing effect of turbulent flows is most intuitively characterized by the growth of the distance between two tracer particles $r$:
\begin{equation} \label{cf3}
 \frac d{dt}\left<\ln r\right>\propto \frac d{dt}\left<[\ln r]^2\right>\propto r^{\xi-2},
\end{equation}
c.f.\ \cite{JK}.
So, with a proper time unit, the distance doubling time is estimated as $\tau\approx r^{2-\xi}$; 
the Kolmogorov scaling $\tau\approx r^{2/3}$ is matched with  $\xi=4/3$.  For the sake of simplicity, 
we consider two-dimensional (2D) geometry (generalization to the 3D geometry is straightforward).

The formation of tracer discontinuities can be qualitatively explained as follows. 
First, we decompose the velocity field into components of different characteristic space-scale of size $a$, $\bm v_a (\bm r , t) =\int_{a\le |\bm k| < 2a} \bm v(\bm k, t) e^{i\bm k \bm r}d\bm k$, where 
$\bm v(\bm k, t)$ is the Fourier component of $\bm v(\bm r, t)$. 
The characteristic time-scale $\tau_a$ 
is defined as the time needed for the field  $\bm v_a (\bm r , t)$ to transport a tracer particle to a distance of the order of $a$;
according to Eq.~(\ref{cf3}),  $\tau_a \approx a^{2-\xi}$.

\begin{figure}
\includegraphics{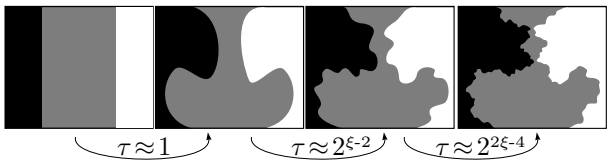}
\caption{Simplified scheme of the formation of tracer discontinuities. Characteristic time of eddies of size $a$ scales as $\tau\approx a^{2-\xi}$.  Due to the combined effect of large and small eddies, 
low- and high-density regions (black and white, respectively) are brought into contact within a finite time.}
\end{figure}
Suppose that initially ($t=0$), there is a constant tracer gradient: 
$\theta(\bm r, 0)\equiv x$. In Fig.~1, the regions $\theta \le 0$, $0< \theta\le 1$, and $1< \theta$ are marked by black, gray, and white, respectively. 
Minimal distance between the isolines $\theta=0$ and $\theta=1$ will start decreasing
(they are turbulently stretched, and the surface are between them is conserved).
Initially, this approaching is dominated by the largest eddies fitting between these lines, i.e.\ the eddies of approximately unit diameter, characterized by $\tau_1\approx 1$.
Then, after a unit time, the distance between some segments of the isolines will be decreased 
approximately by a factor of two. From now on, the distance decreasing rate will be dominated by twice smaller 
eddies, and the characteristic time-scale is reduced by a factor of $2^{2-\xi}$. The process will continue {\em ad infinitum}, leading to the 
contact of segments within finite time (the characteristic time scales form a geometric progression). 

We aim to construct a model, which mimics the evolution of the 
tracer density profile along the $x$-axes (any 1D cross-section). 
To begin with, we consider only the effect of 
an ``$a$-flow''' $\bm v_a (\bm r , t)$ (this corresponds to observing the initial tracer field evolution with 
a spatial resolution $a$: smaller vortices are not resolved, larger ones are slower and require a longer observation period).
In incompressible velocity fields, exponential growth of scalar density gradients is due to exponential stretching of fluid elements, caused by  stretching-folding 
motion of the fluid (c.f.\ \cite{JK0}). Such a stretching-folding  motion is provided by a simple shear flow, as 
depicted  in Fig.~2. Now, consider the tracer density profile along $x$-axes in Fig~2: initially monotonous curve $\theta (x)$ is replaced by a new 
profile $\theta^\prime (x)$ with a ``kink''. The kink emerges because a  descending segment is substituted 
by the sequence of descending, ascending, and descending segments. In the idealized version, all these curve segments are mirror images of each other and
the result of a 3-fold ``compression'' of the initial curve segment along the $x$-axes. Such a mapping ${\cal M}_{a,c}: \theta (x)\to\theta^\prime (x)$ is
represented in Fig.~3(A); here, $a$ denotes the size of the vortex and $c$ --- its middlepoint.
\begin{figure}
\includegraphics{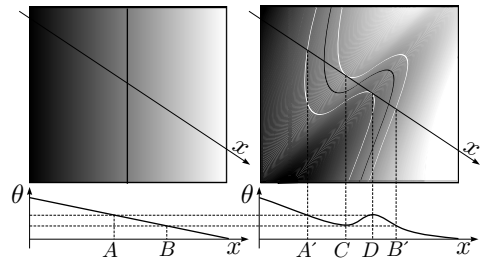}
\caption{The effect of a shear flow on the tracer density profile along the $x$-axes. 
Suppose that initially, the tracer density gradient is constant (left image). Initially straight 
isolines evolve into s-like curves (black and white curves in right image; $C$ and $D$ are their touching points with the $x$-axes).
The segment $AB$ of the initial profile $\theta (x)$ evolves into a ``kink'' of the final profile $\theta^\prime (x)$, consisting of 
descending, ascending, and descending segments $A^\prime C$, $CD$, and $DB^\prime$, respectively.}
\end{figure}
\begin{figure}
\includegraphics{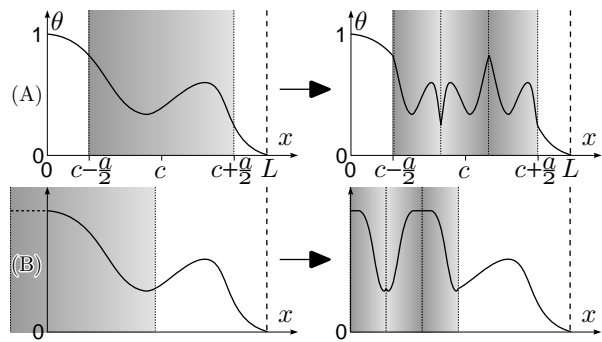}
\caption{Mapping ${\cal M}_{a,c}$ modeling the effect of a single vortex of size $a$ on the tracer profile $\theta (x)$. The vortex may be far from the vessel boundary [Case (A)], 
or it may be close to the boundary, at which a fixed tracer density value $\theta|_{x=0} \equiv 1$ is kept [Case (B)].}
\end{figure}

Our model for the tracer turbulence is iterative application of the mapping ${\cal M}_{a_t,c_t}$ to some initial profile $\theta_0(x)$ with random values of the parameters $a_n$ and $c_n$:
$\theta_{t+1}(x)={\cal M}_{a_t,c_t}[\theta_{t}(x)]$; note that $t$ plays the role of (discrete) time.
In order to match statistically homogeneous turbulence, the probability distribution function (PDF) of this mapping over the parameter $c$ needs to be homogeneous, and 
PDF over $a$ needs to match the stretching statistics of the velocity field (\ref{cf3}). 
Therefore, the waiting time $T_a$ between two subsequent mappings of size $a$ at $x=c$ (i.e.\ satisfying the conditions $a_t \in [a,2a]$ and $c \in [c_t-a_t/2,c_t+a_t/2]$) needs to scale as $T_a\approx a^{2-\xi}$.

We need also to address the issue of the boundary conditions. Initial conditions in the form $\theta(t=0,\bm{r})$ are modelled by the initial profile  $\theta_0(x)$. 
The situation when there is no tracer flux through the container boundaries can be modelled by periodic boundary condition $\theta_t(x+1)\equiv\theta(x)$. The effect of forcing  in Eq.~(\ref{ode}) is 
mimicked by additional iterations $\theta_{t+1}(x)=\theta_{t}(x)+f(t,x)$. If there is a boundary with a fixed tracer density $\theta|{x=0}=1$, we need to incorporate a mechanism of tracer influx at that density. 
For real turbulent flows, that influx is provided by these vortices, which are  in contact with the wall, and is proportional to the vortex size. A convenient way to match such a process is represented in Fig.~3(B).
If the outer edge of the mapping  ${\cal M}_{a_t,c_t}$ happens to be beyond the container boundary $x=0$ (i.e.\ $c_t<a_t/2$), the profile $\theta_t(x)$ 
is extended to the region $x<0$ with the value, kept at the boundary. Then, the mapping can be applied in the same way as described before [with a single modification: ``compression'' factor
is increased from 3 to $3(\frac 32 -c_t /a_t)$, so that the entire ``vortex'' will fit inside the region $x>0$].

Second, we need to discuss the effect of seed diffusivity. For any non-zero diffusivity $\kappa$, the diffusion smoothes the tracer density fluctuations at a microscale $\delta$, for 
which the effective Peclet' number $P_\delta\approx 1$. From the equality of diffusion and mixing times, $\tau_\delta\approx\delta^{2-\xi}\approx \delta^2/\kappa$, we obtain $\delta\approx \kappa^{1/\xi}$.
In order to take into account such a smoothing, the mapping  ${\cal M}_{a_t,c_t}$ is modified so that apart from the effect depicted in Fig.~3, it includes also averaging over 
a sliding window of width $\delta$. 

For numerical simulations, $\delta$ serves as a natural discretization step: tracer density profile is stored as an array $\theta_{i,t}\equiv\theta_t(i\delta)$, $i=1,2,\ldots N$ with $N=\kappa^{-1/\xi}$. 
Then, the iteration formula is
$\theta_{i,t+1}=\frac 13\sum_{|k-j|\le 1}\theta_{k,t}$, where $j=\tilde c_t-3(i-\tilde c_t)$, if $i-\tilde c_t\in [-\frac 13\tilde a_t, \frac 13\tilde a_t]$, and 
$j=\tilde c_t+3(i-\tilde c_t\pm \frac 13\tilde a_t)$, if $i-\tilde c_t\pm \frac 13\tilde a_t\in [-\frac 13 \tilde a_t, \frac 13 \tilde a_t]$. Here, 
$\tilde a_t= a_t/\delta$ and $\tilde c_t=c_t/\delta$ are the discretized mapping parameters.

{\em II.\ Scaling analysis of the model.} 
Our scaling analysis is based on the probability density function (PDF) $f_a(\Delta_a)$ of the difference $\Delta_a(x)=|\bar \theta_a (x+\frac a2) - \bar \theta_a (x-\frac a2)|$ 
between the mean values of the tracer densities for neighboring segments of length $a$. 
Here, the local average is defined as 
$\bar \theta_a(x) = \frac 1a\int_{x-\frac a2}^{x+\frac a2}\theta(y)dy$. 

\begin{figure}
\includegraphics{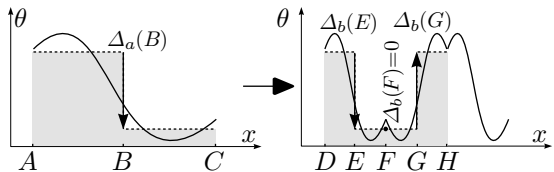}
\caption{As a result of the mapping ${\cal M}_{2a,B}$, the old value of the mean density difference $\Delta_a(B)$ defines the possible range of new values  at trice smaller scale $b=a/3$:
the smallest value is $\Delta_b(F)=0$, and the largest one $\Delta_b(E)=\Delta_b(G)=\Delta_a(B)$.}
\end{figure}
To begin with, we consider, what will happen, if segments $AB$ and $BC$, characterized by mean densities $\bar\theta_{AB}$ and $\bar\theta_{BC}$, 
are transformed by a mapping, see Fig.~4. The segment $AB$ is transformed into three trice smaller segments, two of which are marked as $DE$ and $GH$;
the mean densities for these segments are equal to that of the segment $AB$. The same applies to the segments $BC$, $EF$, and $FG$.
So, the density drop between the segments $DE$ and $EF$ equals to that of between the segments $AB$ and $BC$, i.e. to $\Delta_a(B)$. However, there is no density drop between the 
segments $EF$ and $FG$. Apparently, the density drop $\Delta_{a/3}(x)$ takes all the intermediate values between
$\Delta_a(B)$ and 0, as $x$ moves from $E$ to $F$; the dependence on $x$ is approximately linear (at least in the neighborhood of $F$).
Consequently, as a result of the mapping, the probability $f_a(\Delta_a)d\Delta_a$, associated with the density drop $\Delta_a$, contributes to the PDF $f_{a/3}$ evenly over the range of 
values $\Delta_{a/3}\in[0,\Delta_a]$. This mechanism relates all the values of $f_b(\Delta_b)$ with $b=\frac a3$ to the values of $f_{a}(\Delta_{a})$ 
[because mappings are continuously being applied to the profile $\theta(x)$];
mathematically,
\begin{equation} \label{f1}
 f_b(\Delta)=\int_\Delta^1f_a(\Delta^\prime)\frac {d\Delta^\prime}{\Delta^\prime}
\end{equation}
(assuming that the maximal value of $\Delta$ is 1). 

\begin{figure}
\includegraphics{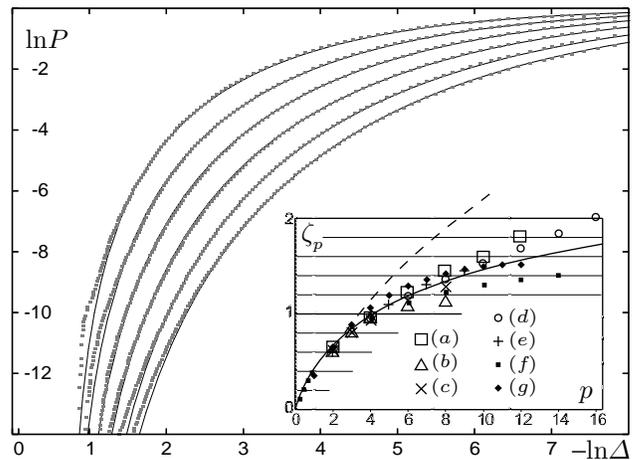}
\caption{Theoretical curve family of the cumulative density difference probabilities $P=\int_{\Delta^\prime\ge \Delta} f_a(\Delta^\prime)d\Delta^\prime$ is calculated according to Eq.~(\ref{f2})
for $\xi=\frac 43$ and $a=\delta, 4\delta, \ldots, 4^5\delta$ (solid curves). The grey dots indicate the corresponding data-series of our simulations. Insert: the 
theoretical $\zeta_p$-curve (solid line; $\xi=\frac 43$) is compared with various experiments and Navier-Stokes numerical simulations; the dashed line corresponds to the Kraichnan formula \cite{Kraich}. 
The data-series are from the following references:
($a$) --- \cite{Antonia}, ($b$) --- \cite{Meneveau}
($c$) --- \cite{Ruiz}, ($d$) --- \cite{Mydlarski}, ($e$) ---  \cite{Chen}, ($f$) --- \cite{Celani}, ($g$) --- \cite{Moisy}}
\end{figure}
It should be emphasized that Eq.~(\ref{f1}) is obtained, using two implicit assumptions. {\em (i)} We do not consider the effect of those mappings, the size of which is either significantly 
larger or smaller than $a$ and $b$. It can be argued that the effect of smaller size mappings is insignificant at our scale, because they preserve the average density $\bar\theta_b$ 
(this is true, if the mapping falls entirely into the segment; if it falls at the edge, $\bar\theta_b$ will be changed, but the change remains relatively small). However, for very small values of $\xi<\xi_0$, 
when small vortices are much more frequent than the large ones, this assumption will no longer be valid. {\em (ii)} We can neglect the effect of larger vortices.
This is actually not true: larger-size mappings compress the profile without reducing the density drop $\Delta$ via the process depicted in Fig.~4.
Such a process corresponds to a direct transfer  $f_a(\Delta)\to f_{a/3}(\Delta)$, without the convolution in Eq~(\ref{f1}). So, in average, the profile will be compressed more
than trice, before entering the convolution stage. 
Hence, the effect of larger vortices can be taken into account by using an effective, somewhat increased compression factor $k=a/b>3$.

Bearing in mind boundary conditions $\theta(0,t)\equiv 1$ and $\theta(1,t)\equiv 0$, it is reasonable to assume that
$f_1(\Delta)\equiv 1$.  Then, direct integration results in $f_a(\Delta_a)=|\ln(\Delta_a)|^n/n!$, where $n=-\log_ka$ is an effective number of iterations.
Now we can easily calculate the structure function scaling exponents $\zeta_p$. Indeed, we expect that $\int f_a(\Delta)\Delta^pd\Delta\propto a^{\zeta_p}$; the integral is easily taken, resulting in
$\zeta_p=\log_k(p+1)$. Comparing this expression with the classical result $\zeta_2=2-\xi$ \cite{Obukhov} (which is valid both for tracer turbulence, and for our 1D model), we obtain $k=3^{1/(2-\xi)}$.
This equality allows us to rewrite the expressions of $f_a$ and $\xi_p$ as 
\begin{equation} \label{f2}
\begin{array}{l}
f_a(\Delta)=|\ln(\Delta)|^nn!^{-1},\; n=-(2-\xi)\log_3a,\\
\zeta_p=(2-\xi)\log_3(p+1).
\end{array}
\end{equation}

\begin{figure}
\includegraphics{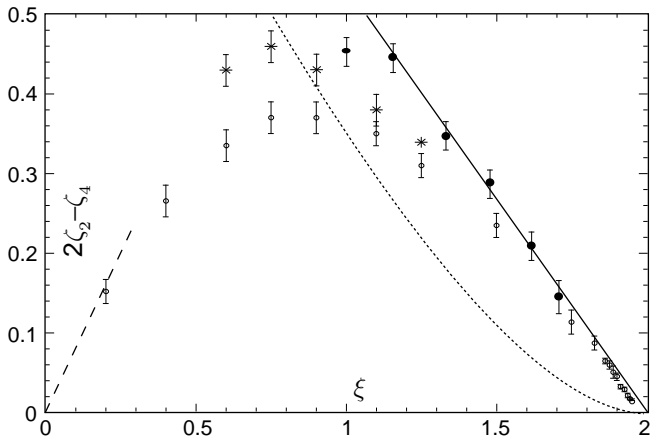}
\caption{The dependence of the second order anomalous scaling exponent $2\zeta_2-\zeta_4$ on the smoothness exponent $\xi$. Solid line corresponds to the theoretical curve $(2-\xi)\log_3\frac 95$,
dotted line --- to the Kraichnan formula \cite{Kraich}, dashed line --- to the perturbation theory \cite{Gaw}, open circles and stars --- to the 
numerical results of 2D and 3D Lagrangian simulations \cite{Frisch}, and filled circles --- to the simulations with our 1D model.}
\end{figure}

Now, let us recall that we expected $k\ge3$; this inequality is not satisfied for $\xi<1$. So, we can conclude that $\xi_0=1$, i.e.\ for $\xi<1$, the assumption {\em(i)} is not satisfied. Note that
the result $\xi_0=1$ is directly applicable only to our 1D model, when all the compression factors are equal to 3; in the case of real 2D or 3D turbulence, the effective compression factors may take different
values and hence, the critical value  $\xi_0$ may deviate from 1.

We have implemented the above described model [with boundary conditions  $\theta(0,t)\equiv 1$ and $\theta(1,t)\equiv 0$] numerically for several values of $\xi$. 
The array length was taken equal to $N=10^6$ (recall that this parameter plays the role of the ratio of the tank width and diffusion scale).
For each value of $\xi$, the observation time of the evolution of the density field $\theta_{t,i}$ was long enough to include at least $10^6$ full decorrelations 
(i.e.\ at least $200$ occurrences of the largest-sized mappings with $a\ge \frac 12$). Simulation results are presented in  Fig.~5 and Fig.~6. 
The small mismatch between the curves and data points in left-hand-side of Fig.~5 can be explained by finite-size effects and somewhat purer statistics of extreme events.
The reason of the departure of the theoretical curve from the simulation data for $\xi\alt 1$ has been already discussed.

In conclusion, our 1D model and analytical results explain, with a reasonable accuracy, the results of previous experiments and simulations. Eq.~(\ref{f2}) predicts that there is no 
saturation of the exponents $\zeta_p$ (for $p\to\infty$). While some experiments have reported such a saturation (c.f. \cite{Celani}), the others have not (c.f.\ \cite{Mydlarski}; the inconsistent 
results can be explained by pure statistics of extreme events (very large density differences). The possibility to extend our approach to passive and active vectors 
(kinematic dynamo and hydrodynamic turbulence problems) will be the scope of further studies.


The support of Estonian Science Foundation grant No.~6121 is acknowledged.

\end{document}